          [2001/04/25 1.1 (PWD)]
\documentclass[a4paper,twocolumn]{esapub2005} 
\usepackage{times}
\usepackage{natbib}
\usepackage{graphicx}
\usepackage{amssymb}

\title{RELATIVISTIC VERSUS NEWTONIAN FRAMES: EMISSION COORDINATES}
\author{J.-F. Pascual-S\'anchez}
\author{A. San Miguel}
\author{F. Vicente}
\affil{Dept. Matem\'atica Aplicada, Facultad de Ciencias,
 Universidad de Valladolid, Valladolid, 47005, Spain, E.U.}

\begin{document}

\keywords{null emission coordinates, location systems, causal class,
positioning system, gravimetry}

\maketitle

\begin{abstract}
 Only a causal class  among the 199
Lorentzian ones, which do not exists in the Newtonian spacetime,
is pri-vileged to construct a generic, gravity free and immediate (non retarded) relativistic positioning system.
 This is the causal class of the null emission coordinates. Emission coordinates are  defined
and generated by four emitters broadcasting their
proper times. The emission coordinates are covariant (frame independent)
and hence valid for any user. Any observer can obtain the values
of his (her) null emission coordinates from the emitters
 which provide him his (her)
trajectory.
\end{abstract}

\section{Introduction}

Globally, the current situation in the Global
 Navigation Satellite Systems (GNSS) is almost analogous to the fo-llowing one:
imagine that a
century after Kepler, the astronomers were still using Kepler's laws as algorithms
to correct the Ptolemaic
epicycles by means of ``Keplerian effects". Similarly, a century after Einstein,
one still uses the Newtonian
theory and corrects it by ``relativistic or Einsteinian effects" instead of starting
 with Einstein's gravitational
theory from the beginning.

To show this, we will focus on the essential differences between an old
newtonian plus ``relativistic corrections" coming from the post-Newtonian framework, as in the current
 operating systems which use only the usual class of Newtonian frames, and the new fully relativistic framework which use a new class of relativistic frame: null emission coordinates. Note that there are not ``relativistic corrections" in relativity, as they are not ``Newtonian corrections" in Newtonian theory.

 At present,  the GNSS  functioning  as  global positioning systems,
are the GPS and the GLONASS. In general, the satellites (SVs) of the GNSS are affected by Relativity
in three different ways: in the equations of motion, in the signal
propagation and in the beat rate of the satellite clocks.
We will only briefly comment on the clock effects because they are only the
measurable ones in the present GNSS and in the future Galileo.

 Among the relativistic effects on the rate of the satellite clocks with a time accuracy
 of nanoseconds and $10^{-12}$ of frequency accuracy,
 the most important ones (to first post-newtonian order
$1/c^2$) are: the Einstein effect or gravitational frequency blue shift
of the atomic clocks of the satellites (Equivalence principle of
General Re-lativity) with respect to Earth bound clocks due to their position in the Earth gravitational field, time dilation or Doppler shift of second order due to
the speed of the satellites (Special Relativity) and the kinematical Sagnac effect due to
the rotation of the Earth (Special Relati-vity), see Refs.
\citep{ash} and  \citep{pascual} for  reviews. If they were
not corrected by imposing an offset, the GNSS would not be operative after few
minutes.

However,  with the present and future more
accurate clocks (pico and even femtosecond), it would be nece-ssary in the Newtonian framework   to consider
 other ``relativistic corrections" at post-post-Newtonian order as well as  metric spatial curvature effects, tidal effects or delay effects of gravity in the light propagation as
 the Shapiro time delay.

In this situation, it can be wondered if it
would not be more convenient to change the present Newtonian
framework to an exact formulation in full General Relativi-ty. This
would imply to abandon the classical post-newtonian framework for the description of GNSS.
The root of this radical change is the consideration of a new 4D proper relativistic frame
(emission coordinates) ins-tead of the usual Newtonian frame, which uses   3D spatial
reference systems, such as the ECI (Earth Centered Inertial system) or the ECEF (Earth Centered Earth Fixed system), and a time
reference (GPS time), separately.

Emission coordinates were firstly introduced by B. Coll  in a pioneering proposal
presented at the ERE–2000 Spa-nish Relativity Meeting and published in \citep{coll1}.
To discuss and understand the meaning of the null emission coordinate system is necessary
 to introduce previously some new definitions, as such location systems or causal classification of frames, and
  mathematical physics tools, mainly geometrical. These new definitions and tools provide a
  clear way to understand the differences among the special subclass of Newtonian frames and
  the general class of relativistic frames.

\section{Location systems}
Location systems are physical realizations of 4D coordinate systems. Hence there is a
differentiation of a  coordinate system as a mathematical object from its realization
through physical objects and protocols. A location system is thus a
precise protocol on a  set of physical fields allowing to materialize a coordinate
system. However, different physical protocols, involving different physical fields,
may be given for a unique mathematical coordinate system.

A location system must include the protocols for the physical construction of
the coordinate lines, coordinate surfaces or coordinate hypersurfaces of specific causal orientations of the coordinate
system that it  realizes. Thus, for instance, these coordinate elements may
be realized by means of clocks for timelike lines, laser pulses for
null lines, synchronized inextensible threads for spacelike lines,
 laser beams or inextensible threads for time like surfaces and light-front surfaces
 for null hypersurfaces. The different protocols involved in the construction of location systems give rise to
coordinate e-lements (lines, surfaces and hypersurfaces) of different causal orientations,
i.e., they realize coordinate systems of different causal nature.

\subsection{Reference systems}
 Location systems are of two different types: reference systems and positioning systems. The
first ones are 4D reference systems which allow one observer, considered at the
origin, to assign four coordinates to the events of its neighborhood by means of  electromagnetic
signals. In relativity due to the finite speed of the transmission of information, this assignment is retarded with a time delay.

A paradigmatic reference system in relativity is the {\sl radar system} which is based in the Poincar\'e protocol
of synchronization which uses two-way light signals from the observer to the events to be coordinated. Unfortunately, the radar system suffers from the bad property of being constructed from a retarded protocol due to the finite speed  of the transmission of information.

\subsection{Positioning systems}
The second kind of location systems are 4D positioning systems, which allow to every event of a given domain to know its proper coordinates in an immediate or instantaneous way without delay. In addition to be immediate, the positioning systems must verify other two conditions, they must be generic and free of gravity.  A positioning system is generic, if it can be constructed in any spacetime and, it is free of gravity, if the knowledge of the gravitational field is not necessary to construct it.
Re-ference systems privilege one specific observer among all others, whereas in positioning systems no observers are necessary at all and hence there is no necessity of any synchronization
procedure between different observers.

In relativity, a (retarded) reference system can be constructed starting from an (immediate) positioning system, it is sufficient that each event sends its coordinates to the observer
at the origin of the reference system, but not the other way around. In contrast, in Newtonian theory, 3D
reference and positioning systems are interchangeable and as the velocity of transmission of information is infinite, the Newtonian reference systems
are not retarded but immediate.
The reference and positioning systems defined here are 4-dimensional objects, including
time location.

 \section{causal classification of frames}
 In the Lorentzian spacetime of general relativity, directions and planes or hyperplanes of directions at any event are said to be
spacelike, lightlike  (or  null or isotropic) or timelike oriented if they are respectively exterior, tangent or secant to the light-cone of this event. These causal orientations can be extended in a  natural way to vectors, covectors and volume forms on these sets of directions. Thus, every one of the vectors $e_A$ of a frame of the tangent space $\{e_A\}$\, $(A = 1,...,4)$ has a particular causal orientation ${\rm c}_A\,.$

 However, the causal orientations ${\rm C}_{AB}$\, $(A<B)$ of the  six different  associated or adjoint  planes
$\Pi(e_A,e_B)$ of the frame $\{e_A\}$\, are {\em not} determined by the  specific causal orientations ${\rm c}_A$ of the vectors of the frame. For instance, the plane associated to two spacelike vectors may have any causal
orientation. So, in general, the causal characters ${\rm c}_A$ and ${\rm C}_{AB}$ are independent.
Moreover, in order to give a complete description of the causal properties of the frames, one needs also to
specify the causal orientations $\it{c}_A$ of the four covectors or 1-forms $\theta^A$ giving the dual coframe $\{\theta^A\}$, i.e. $\theta^A (e_B) = \delta^A _B$.
Following \citep{coll2}, the best  way to visualize and characterize
a spacetime coordinate system is to start from four families of
coordinate 3-surfaces, then, their mutual intersections give six
families of coordinate 2-surfaces and four congruences of coordinate lines.

 Alternatively,  one can use the related covectors or 1-forms
$\theta^A$, instead of the 3-surfaces, and
the vectors of a coordinate tangent frame $\{e_A\}$\,,  instead of four congruen-ces of coordinate lines which are
their integral curves. The covector $\theta^A$  is
 timelike (resp. spacelike) iff the 3-plane $\Pi(e_B, e_C, e_D)$ generated by the
 three vectors $\{e_B\}_{B\neq A}$ is spacelike (resp.
 timelike). This applies for both Newtonian and Lorentzian
 spacetimes. In addition, for the latter, the covector $\theta^A$ is lightlike (or null) iff the 3-plane generated by $\{e_B\}_{B\neq
 A}$ is lightlike (or null). Thus, to specify the causal orientations of hyperplanes is not necessary because is redundant with the causal orientation of the covectors.

In this way, for a specific domain of a Lorentzian or Newtonian spacetime, each frame $\{e_A\}$\, is fully characterized by its causal class. The causal class
of a frame is the set of all the frames that have same causal signature, which is defined by a set of 14 causal characters:
\begin{equation}\label{22}
\{{\rm{c_1\, c_2\, c_3\, c_4, \,C_{12}\, C_{13}\, C_{14}\, C_{23}\,
C_{24}\, C_{34}}},\,{\it c_1\, c_2\, c_3\, c_4}\},
\end{equation}
As notation for the causal characters, we will use lower case roman
types ({\rm{s,\,t,\,l}}) to represent the causal character of vectors
(resp. spacelike, timelike, lightlike), and capital types (S,\,T,\,L)
and lower case italic types ({\it s,\,t,\,l}) to denote the causal
character of 2-planes and covectors, respectively.

\subsection{Relativistic frames}
This new degree of freedom (lightlike) in the causal
 character, which is proper of Lorentzian relativistic spacetimes
 but which does not exist in Newtonian spacetimes,  allows to obtain (see \cite{coll2}), as it has been commented
  in the abstract, the following theorem: {\sl In a
    relativistic 4-dimensional Lorentzian spacetime, there exists 199,
   and only 199, causal classes of frames.} These 199
    causal classes have been completely classified.

We shall see that among the 199 Lorentzian causal classes, only one is privileged to construct a generic,
gra-vity free and immediate positioning system.

   The notion of causal class extends naturally to the set
of coordinate lines of the coordinate system and so, to the coordinate system itself.
By definition, the causal class of a coordinate system $\{x^\alpha\}_{\alpha=1}^4$ in a domain is the causal class
$\{\rm c_\alpha, \,\rm C_{\alpha\beta},\, \it{c}_\alpha\}$ of its associated {\sl natural} frame at the events of the
domain. In relativity, a specific causal class, among the 199 ones, can be
assigned to any of the different  coordinate systems used in all the
solutions of the Einstein equations. However, for the same
coordinate system and the same solution, the causal class can change
depending on the region of the spacetime considered and the coordinate system in this case is said to be inhomogeneous.

In fact, see \cite{mor}, in any spacetime every coordinate $x^\alpha$ plays two extreme roles: that of a
hypersurface for every constant value $x^\alpha=const$, of gradient $dx^\alpha$, and that of a coordinate line of tangent vector $\partial_\alpha$, when the other
coordinates remain constant. This simple fact shows that, in spite of the
historical custom of associating to a coordinate a causal orientation, saying that {\sl it is} timelike,
lightlike or spacelike, {\sl this appellation is not generically coherent.} Causal orientations are generically
associated with directions of geometric objects, but not with spacetime coordinates
 associated to them. In the case of a coordinate $x^\alpha$, this generic incoherence appears because
its two natural variations in the coordinate system, $dx^\alpha$ and $\partial_\alpha$, have  generically
different causal orientations.  {\sl Only when both causal orientations coincide}, it is possible to
extend to the coordinate $x^\alpha$ itself the character of the  common causal orientation of its two mentioned variations.

\subsection{Newtonian frames}
The differences in the geometric description of Lorentzian and Newtonian frames come
from the causal structure induced by the different metric descriptions of Lorentzian and Newtonian spacetimes.
The main difference comes essentially from the absence of the lightlike character in the Newtonian case.
In relativity, the spacetime metric is non-degenerate and defines a one-to-one correspondence between vectors and covectors at the tangent and cotangent space of every event.

In contrast, in a Newtonian space-time no non-degenerate metric structure exists and one have two
 different me-trics, see \cite{Trautman}. This degenerate metric
structure is given by a rank one {\sl covariant   time metric} $T=dt^2$ and an orthogonal rank three
{\sl contravariant   space metric} $\gamma$. In the time metric appears  $t$ which is a absolute time
scale and the hypersurfaces $t = const$ constitute the instantaneous or simultaneity spaces.
 A vector $e$ is  spacelike if it is instantaneous, i.e. if $dt(e)= 0$. Otherwise, it is is timelike.
So, it is clear that a frame can have at most three spacelike vectors so there only exist four causal
types of Newtonian frame bases, namely: $\{\rm{t s s s} \},$ $ \{\rm{t t s s} \},$ $ \{\rm{t t t s} \},$ $ \{\rm{t t t
t} \}.$

Correspondingly, a covector $\theta \neq 0$ is  timelike if it has no instantaneous
part with respect to the contravariant space metric $\gamma$, i.e. if $\gamma(\theta)=0$ and it is necessarily of the form $\theta = N \, dt$ with $N \neq 0$, being {\sl future} (resp. {\sl past}) {\sl oriented} if $N>0$ (resp. $N<0$). Otherwise, the covector $\theta$ is  spacelike. Thus, attending to the causal orientation of their covectors, there only exist two causal types of Newtonian coframes bases: $\{{\it t s s s}\}, \{{\it s s s s}\}$.

In summary, it can be shown (see \cite{mor}) that  one has the following implications valid only for Newtonian frames:
$    \{{\rm c_A}\}\Rightarrow \{{\rm C_{AB}}, {\it c_A}\}  \, ,
    \quad \{{\rm C_{AB}}\}\Rightarrow \{{\it c_A}\}  ,
$ but  $  \{{\rm C_{AB}}\} \nRightarrow \{{\rm c_A}\} \, ,
    \quad \{{\it c_A}\} \nRightarrow \{{\rm c_A}, {\rm C_{AB}}\} \quad .
$

The simplicity of the  Newtonian causal structure with respect to the Lorentzian one lies in the fact
that the causal type of a Newtonian frame determines completely  its causal class. This is related to the fact
that, in Newtonian space-time, any set of spacelike vectors always generates a spacelike subspace. As a
consequence, the number of causally different Newtonian classes of frames is equal to the dimension of the spacetime.
Hence, see \cite{mor}, in {\sl the 4-dimensional Newtonian spacetime there exist four, and only four,
causal classes of frames}. They are:
 $\{{\rm{t s s s, \,T T T SSS}},\,{\it t s s s }\}$,
 $\{{\rm{t t s s, \,TTTTTS}},\,{\it s s s s}\}$,
 $\{{\rm{t t t s, \,TTTTTT}},\,{\it s s s s}\}$ and
 $\{{\rm{t t t t, \,TTTTTT}},\,{\it  s s s s}\}$.

For instance, the standard spatial coordinates  ECI and ECEF used in the GPS more the GPS time, i.e. those that are locally realized with three rods and one clock, belong to the same causal class
$\{{\rm{t s s s, \,T T T SSS}},\,{\it t s s s }\}$, the first one  above. The history of the clock is a timelike coordinate line. The other
coordinate lines are spacelike straight lines tangent to the rods at every time. Also the reference systems adopted by the I.A.U. for the Earth and the barycenter of the Solar system as, respectively,  the
3-dimensional International Terrestrial Reference System (ITRF),
which is also an  ECEF, plus the International Atomic Time
(TAI) and the 3-dimensional International Celestial Reference Frame (ICRF) plus the TCB time belong to this usual causal Newtonian class.

\section{Relativistic positioning}
\subsection{Coll positioning system}
As it has been commented above, among the 199 Lorentzian causal classes, in which the four Newtonian ones
are included, only one is privileged
to construct a generic (valid for a wide class of spacetimes),
gravity free (the previous knowledge of the gravitational field is
not necessary) and immediate relativistic positioning system.  This is the causal
class $\{{\rm{s\,s\,s\,s,S\,S\,S\,S\,S\,S}}, l\,l\,l\,l \}$ of the
Coll homogeneous coordinate system \cite{coll1,pascual,coll2}. In this causal class
the {\sl null emission coordinates} of the Coll positioning system are
included. These emission coordinates have been also studied in
\cite{rovelli,Blagojevic,coll3} in the special case of a flat Minkowski
spacetime without gravity.

The coordinate system of this causal class is always homogeneous and it has associated four
families of null 3-surfaces or equivalently a real non-orthogonal null coframe, whose mutual
intersections give six families of spacelike 2-surfaces and four
congruences of spacelike lines. Such a coordinate system does not exist in a Newtonian space-time where the
light travels at infinite speed.
\begin{figure}
\centering
 \includegraphics[width=3.2in]{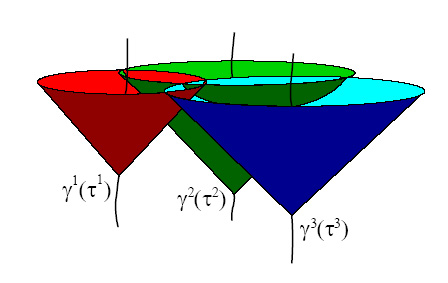}
\caption{{\small  Relativistic emission coordinates:
intersection of the four future light cones of the SVs with  the
past light cone of a receiver. In the Figure only 3 light cones of
the SVs are drawn in a Lorentzian spacetime of 3 dimensions.}\label{fig2}}
\end{figure}
One satellite clock broadcasting its proper time is des-cribed
in the spacetime by a timelike line $\gamma^A(\tau^A)$ in which each event of proper time $\tau^A$ is the
vertex of a future light cone. The set of these light cones of a
emitter constitutes a one-parameter (proper time) family of null
hypersurfaces. So, four satellite clocks broadcasting their proper times
determine four one-parameter families of  lightlike
3-surfaces (future light cones), see Figure ˜\ref{fig2}. Thus, the Coll positioning system makes use of the ma-thematical fact that four future
light cones generically intersect in an unique event, which is just the spacetime position of the receiver or user.

\begin{figure}
\centering
 \includegraphics[width=0.6\linewidth]{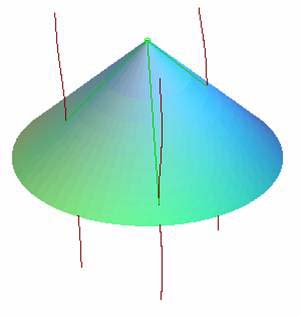}
 \caption{{\small  Past light cone of an event in  3 dimensions,
the proper time parameterized paths of 3 SVs (in violet) and the
lightlike geodesics (in green) followed by the signal from each
sate-llite to a event of the trajectory of a receiver.}\label{fig3}}
\end{figure}

In this  relativistic positioning
system, any receiver or  user at any event in a given spacetime region can know
its proper coordinates. The four proper times of four sate-llites
 $ (\{\tau^A \};\, A = 1,2,3,4$) read at an event by a receiver or user
constitute the null (or light) proper emission coordinates or user
positioning data of this event, with respect to four SVs, see Figure ˜\ref{fig3}.
These four numbers or pa-rameters can be understood as the ``distances" between
the reception event and the four satellites.

In  a certain domain
$\Omega\subset\mathbb{R}^4$ of the grid of parameters $ \{\tau^A
\} $, any user receiving continuously his null emission
coordinates from four satellites may know his trajectory in the grid of parameters. If the observer has his own
clock, with proper time denoted by $\sigma$, then he can know his trajectory with proper time parametrization, $\tau^A=\tau^A(\sigma)$, and  his four-velocity,
$u^A(\sigma)= d \tau^A/d\sigma$.

For positioning out a GNSS constellation, i.e. for interplanetary missions in the Solar system,   a ``pulsar"
Coll relativistic positioning system can be conceived, based on the X-ray signals of four properly selected stable
mi-llisecond pulsars and a conventional origin of the emi-ssion coordinates. On the other hand,  a navigation  project called XNAV (based in pulsars) is being developed during the last  years  by DARPA and NASA but unfortunately, see \cite{graven}, is based in the same  Newtonian concepts that the GPS or Galileo. However, in this case, it is more complicated because post-post-Newtonian corrections must be implemented.

\subsection{Contravariant metric in emission coordinates}
As the emission coordinates belong to the causal class $\{{\rm{s\,s\,s\,s,S\,S\,S\,S\,S\,S}}, l\,l\,l\,l \}$,
there is not a spacetime asymmetry like in
the standard Newtonian coframe  $(\it{t\,s\,s\,s })$ (one
timelike ``$\it{t}$" and three  spacelike ``$\it{s}$''). In
emission coordinates  obtained from a general real null coframe
  $ (\it{l\,l\,l\,l})=\{d\tau^1, d\tau^2, d\tau^3, d\tau^4\}$, which is neither orthogonal nor
   normalized, the contravariant spacetime metric is symmetric with null diagonal
  elements  and it has the general expression \cite{coll5,pascual}:
\begin{equation}\label{25}
g^{AB}=  {d\tau ^A} \cdot d\tau ^B =\left(\begin{array}{cccc} 0 &g^{12}& g^{13}&g^{14}\\
g^{12}&0& g^{23}&g^{24}\\
g^{13}& g^{23}&0&g^{34}\\
g^{14}& g^{24}&g^{34}&0\end{array}\right),
\end{equation}
where $g^{AB}>0$ for $A\neq B$. Four null covectors can be
linearly dependent although none of them is proportional to
another. To ensure that the four null covectors are linearly
independent and span a 4-dim spacetime, it is sufficient that
$\rm{det}(g^{AB})\neq0$. Finally, this metric has a Lorentzian
signature $(+,-,-,-)$ iff $\rm{det}(g^{AB})<0$.

The expression (\ref{25}) of the metric is observer independent and has six
degrees of freedom. In the terminology of \cite{rovelli}, the proper
 times $\tau^A$ are {\sl partial observables}, while the components of
the metric $g^{AB}$ are {\sl complete obser-vables}, i.e, gauge independent
 or invariant quantities under
diffeomorphisms in the Lorentzian spacetime.

A splitting of this metric can be considered, see \cite{coll5},
changing from the six independent components (ten components minus four gauge degrees
of freedom of coordinate transformations) of $g^{AB}$  to a more
convenient set, which neatly separates two shape parameters depending
only on the direction of the covectors $d\tau^A$ or equivalently depending exclusively on the
 trajectories of the emitters, from other four scaling
parameters depending on the length of the covectors or depending on the proper time
of each satellite.

\subsection{SYPOR project: autolocated positioning system}
SYPOR project is the anagram (in French) of Relativistic
Positioning System project. The basic idea of this project, that was conceived by Coll in \cite{coll6} and also exposed  in
\cite{pascual,pas}, is the following one: A satellite
constellation provided with clocks that interchange its proper time
among them (interlinks) and with Earth receivers, is  a fully relativistic
autonomous or autolocated positioning system. Note that, nowadays,  this procedure of proper time auto
navigation can be  technically fulfilled.

\begin{figure}
\centering
  \includegraphics[width=1.7in]{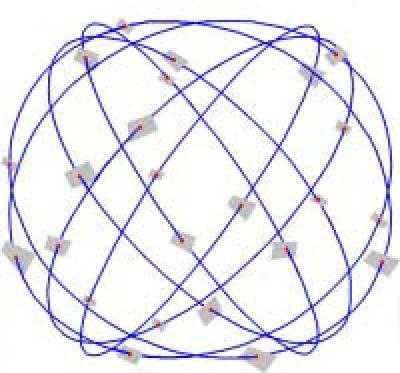}
\caption{{\small  In the SYPOR, the Space and  Control segments
coin-cide with the constellation.}\label{fig4}}
\end{figure}

{\it In the SYPOR, the segment of Control is in the
constellation of satellites}, see Figure ˜\ref{fig4}.
The function of this new Control segment is not to determine the
ephemerides of the satellites with respect to geocentric
coordinates as in the newtonian GNSS, but to determine the
null emission coordinates of the receivers with respect to the
constellation of SVs. Therefore,  the procedure used until now in
the newtonian GNSS  is  inverted.

Let us define properly what
 means autonomous or autolocated.
 Four satellites emitting, without the necessity of a
synchronization convention,  not only its proper times $ \tau^A$,
but also the proper times $ \tau^{AB}$ of the three close
satellites received by the satellite $A$ in $\tau^A$ (in total sixteen emitter positioning data $
\{\tau^A, \tau^{AB}\};\, A \neq B; \,A, B = 1,2,3,4$), constitute
an autolocated positioning system.

In an  autolocated positioning system, the receivers
  can know not only its spacetime path but also
the trajectories of the four satellites in the grid $\mathbb{R}^4$ of emission coordinates.

\section{Gravimetry and positioning}
In General Relativity, the gravitational field is described by the spacetime metric. If this metric is exactly
known a priori, the system just described will constitute an ideal positioning system.  In practice, the actual spacetime me-tric (i.e., the gravitational field) is not exactly known (in the GPS it is supposed to be essentially the Schwarzschild one) and the satellite system itself has to be used to infer it. This problem arises when a satellite system is used for both positioning and gravimetry.

To solve this joint problem, the considered satellites should have more than one clock: they may carry an
accelerometer providing information on the spacetime connection.
Of course, in first approximation the satellites are in free-fall  and consequently have zero acceleration.
 However, we are considering here the realistic case
where the acceleration is nonzero due, for instance, to a small drag in the high atmosphere and this is
measured by the accelerometers. Also, the satellites may have a gradiometer, this would give additional
information on the metric (in fact, on the Riemann tensor of the spacetime). With these data (and
perhaps some additional ones) an optimization procedure could be developed (see \cite{taran}) to obtain  the best observational gravitational field acting ac-tually on the constellation.
The problem of obtaining the spacetime metric is a kind of inverse problem since one wants to recover the spacetime metric from the observed data in the Coll positioning system

\subsection{Two dimensional case}
Coll positioning systems  are yet now quite well developed
for two-dimensional spacetimes, see  \cite{coll3, coll4} were several results have been developed. For instance, the knowledge that the positioning system is stationary  and that the space-time is created by a given mass,
allows to know the  accelerations of the emitters, their mutual radar distances
and the spacetime metric  in null emission coordinates. The
important point for gravimetry is that the Schwarzschild
mass may be substituted by that of the acceleration of one of the emitters.

\subsection{Realistic four dimensional case}
For applications of a autolocated positioning system on or near the Earth's surface, the primary emission coordinates should be related
 to some terrestrial secondary 4-dimensional Newtonian coordinate system. This problem has been solved for a general
configuration of the emitters in flat Minkowski spacetime \cite{coll7} and also for the case of a special configuration of the emitters in a Schwarzschild spacetime \cite{bini}.

 However, in general the known results
for the two dimensional case
are not trivially generalizable for the realistic four dimensional one \cite{coll5}
and much work remains to be done in the future.

\section*{Acknowledgments}

One of the authors (JFPS) is grateful to Bartolom\'e  Coll by communications on this subject along several years and to
Joan Josep Ferrando and Juan Antonio Morales by some friendly discussions.
    The authors are
currently partially supported by  the Spanish Ministerio de
Educaci\'on y Ciencia MEC-FEDER ESP2006-01263 grant.

\end{document}